# Capillary Filling of Star Polymer Melts in Nanopores

*Jianwei Zhang [1], Jinyu Lei [1], Pu Feng [2], George Floudas [3,4,5], Guangzhao Zhang [1,\*], and Jiajia Zhou [6,7,\*]*

1. Faculty of Materials Science and Engineering, South China University of Technology, Guangzhou 510640, China

2. School of Civil Engineering and Transportation, South China University of Technology, Guangzhou 510640, China

3. Max Planck Institute for Polymer Research, 55128 Mainz, Germany

4. Department of Physics, University of Ioannina, 45110 Ioannina, Greece

5. Institute of Materials Science and Computing, University Research Center of Ioannina (URCI), 45110 Ioannina, Greece

6. South China Advanced Institute for Soft Matter Science and Technology, School of Emergent Soft Matter, South China University of Technology, Guangzhou 510640, China

7. Guangdong Provincial Key Laboratory of Functional and Intelligent Hybrid Materials and Devices, South China University of Technology, Guangzhou 510640, China

**ABSTACT:** Topology of polymer profoundly influences on its behavior. However, its effect on imbibition dynamics remains poorly understood. In the present work, capillary filling (during imbibition and following full imbibition) of star polymer melts was investigated by molecular dynamics simulations with a coarse-grained model. The reversal of imbibition dynamics observed for linear-chain systems was also present for star polymers. Star polymers with short arms penetrate slower than the prediction of the Lucas–Washburn equation, while systems with long arms penetrate faster. The radius of gyration increases during confined flow, indicating the orientation and disentanglement of arms. In addition, the higher the functionality of the star polymer, the more entanglement points are retained. Besides, a stiff region near the core segments of the stars is observed, which increases in size with functionality. The proportion of different configurations of the arms (e.g. loops, trains, tails) changes dramatically with the arm length and degree of confinement, but is only influenced by the functionality when the arms are short. Following full imbibition, the different decay rates of the self-correlation function of the core-to-end vector illustrate that arms take a longer time to reach the equilibrium state as the functionality, arm length, and degree of confinement increases, in agreement with recent experimental findings. Furthermore, the star topology induces a stronger effect of adsorption and friction, which becomes more pronounced with increasing functionality.

# I. INTRODUCTION

Recently, star polymers have attracted increasing attention due to their unique molecular topology, linked with several emerging applications.[1,2] One example is their infiltration into narrow pores with size comparable to the polymer size, which is of fundamental interest for application in nanofluidics and membranes.[3-6] Chains under this kind of spatial confinement exhibit nonclassical behavior with respect to their viscosity[7], mobility[8], entanglements[9], etc. The imbibition dynamics of linear chains in this scale also shows deviations from the classical prediction of Lucas-Washburn equation (LWE) [10,11],

$$h(t) = \sqrt{\frac{\gamma R \cos\theta}{2\eta} t} \qquad (1)$$

where $h(t)$ is the filling length of fluid inside the pore and it is a function of wetting time $t$, $\gamma$ is the surface tension, $R$ is the radius of a cylindrical capillary, $\theta$ is the equilibrium contact angle of fluid on the capillary wall, and $\eta$ is the bulk viscosity. Shorter chain polymers exhibit slower imbibition dynamics than predicted by theory, when there is no entanglement or just few entanglements.[12] When the chain length surpasses a certain threshold, the dynamics within the same nanopores reverses and faster-than-prediction imbibition emerges.[13,14] The phenomenon is not only observed inside regular-shaped capillaries, but also within irregular shaped nano-channels in dense packings of nanoparticles. Providing fundamental understanding of this phenomenon is closely linked with applications of porous materials with polymer fillers.[15-17]

Recent works explained the physics behind this phenomenon. The strong interaction between polymer chains and the capillary walls induces a layer of adsorbed chains, which is termed "dead zone".[18] On the macroscopic level, the capillary radius is larger than the size of polymer chains and the thickness of the adsorption layer can be neglected. However, this layer significantly decreases the effective radius of nanopores for small capillaries. Also, the free energy of chains under confinement increases.[18] The combination of these factors results in a slower capillary filling than theoretically predicted. Faster imbibition occurs when polymer chains are well entangled. Under confinement, chains are constrained by other chains and can only move along the "reptation tube", so that the chain mobility is enhanced.[18,19] This effect leads to chain disentanglement during flow, which further results in a reduced effective viscosity. As expected, higher degree of confinement brings about more pronounced effects.[19-21] The change in the interfacial friction or slippage of polymer-wall interface should also be valued in certain cases.[20,22] Besides the imbibition dynamics, a better understanding of chain dynamics in nanopores is also necessary. In previous studies by dielectric spectroscopy, configuration transition of chains from the free one to the fixed one (e.g. loop or train) leads to a piecewise evolution of the dielectric loss curves corresponding to the chain modes.[23] This process also induces extremely slow adsorption kinetics, which is more distinct under higher degree of confinement.[24]

Although the imbibition dynamics have been studied in detail for linear polymers, little is known of how the change in polymer topology influences both the imbibition dynamics and chain dynamics under nanometer confinement. For example, is the mechanism of adsorption sensitive to the polymer architecture? Going one step further, could one employ the imbibition process in nanopores as a means of separating different chain topologies in perfectly miscible polymer blends, e.g., star from linear polymers chains?

Compared with linear chains, star polymers have significantly different dynamics in their bulk state.[25] The simplest star polymer consists of multiple linear chains of the same length grafted onto a core. Even in this case, the basic characteristics of the polymer, such as the chain relaxation or the glass temperature, cannot be described completely by classic models or scaling laws.[26,27] The adsorption kinetics of star polymer at an interface should also be modified by its multi-end topology.[28-30] In addition, much effort has been devoted to find the relation between the fundamental properties of star polymer and its architecture (i.e., functionality $f$ and molecular weight $M$).[31-34] However, it remains to be explored whether these theories can predict the behavior under extreme spatial restriction.

In this work, we employ extensive molecular dynamics simulations on the capillary filling of a series of star polymer melts in nanopores. The whole filling process is divided into two stages: (i) during imbibition, and (ii) after full imbibition. Our research focuses on both the imbibition dynamics and chain dynamics under confinement. During imbibition our aim is two-fold. First, to predict the penetration length of star polymers in comparison to the classical Lucas-Washburn equation, and secondly, to understand the chain behavior (e.g. chain configuration, orientation, disentanglement, and chain mobility) during filling. Following the full imbibition, our objective is to investigate the relaxation of star polymers, as a function of pore size, functionality and arm molecular weight and to evaluate their ability to reach a steady state. Perhaps unsurprising, we find that multi-arm stars behave dynamically different under confinement with respect to linear polymers confined in the same nanopores.

## II. SIMULATION MODEL AND DETAILS

We studied the imbibition dynamics of star polymer melts by performing coarse-grained molecular dynamics simulations. The polymer chains are presented by the bead–spring model and the capillary wall is modeled as ordered spherical beads. Non-bonded interactions between all beads were modeled by the truncated-shifted Lennard-Jones (LJ) potential[35]

$$U_{LJ}(r) = \begin{cases} 4\varepsilon_{LJ}\left[\left(\frac{\sigma}{r_{ij}}\right)^{12} - \left(\frac{\sigma}{r_{ij}}\right)^{6} - \left(\frac{\sigma}{r_{cut}}\right)^{12} + \left(\frac{\sigma}{r_{cut}}\right)^{6}\right] & r_{ij} \leq r_{cut} \\ 0 & r_{ij} > r_{cut} \end{cases} \quad (2)$$

where $\sigma$ is the bead diameter, $\varepsilon_{LJ}$ is the Lennard-Jones interaction parameter, $r_{ij}$ is the distance between the $i$th and $j$th beads, and the cutoff distance $r_{cut}$ was set to 2.5 $\sigma$. For interactions between polymer beads, we set $\varepsilon_{LJ} = 1.4\ k_B T$ and $\sigma = 1.0$, where $k_B$ is the Boltzmann constant, and $T$ is the temperature. The bonds connecting beads into polymer chains were modeled by the finite extension nonlinear elastic (FENE) potential[36]

$$U_{FENE}(r) = -\frac{1}{2}k_{spring}R_{max}^2 \ln\left(1 - \frac{r^2}{R_{max}^2}\right) \quad (3)$$

with the spring constant $k_{spring} = 30\ k_B T/\sigma^2$ and the maximum bond length $R_{max} = 1.5\ \sigma$. Herein, we studied star polymers with three different degrees of polymerization of arms ($N_{arm} = 10, 50,$ and $100$) and four different number of arms ($f = 2, 4, 6,$ and $12$), thus 12 types of polymers in total. The system codes and details are listed in the Table I. The data for the radius of gyration in the table were calculated in the bulk of static equilibrium states (the details will be discussed later).

**TABLE I.** Molecular characteristics of star polymer melts used in the simulation.

| Star polymers | $f$ | $N_{arm}$ | $M$ [a] | $R_g^{(arm)}$ [b] | $R_g^{(star)}$ [c] |
|---|---|---|---|---|---|
| SP10x2 | 2 | 10 | 21 | 1.451 | 2.203 |
| SP10x4 | 4 | 10 | 41 | 1.459 | 2.577 |
| SP10x6 | 6 | 10 | 61 | 1.463 | 2.757 |
| SP10x12 | 12 | 10 | 121 | 1.482 | 3.085 |
| SP50x2 | 2 | 50 | 101 | 3.370 | 4.529 |
| SP50x4 | 4 | 50 | 201 | 3.436 | 5.360 |
| SP50x6 | 6 | 50 | 301 | 3.465 | 5.530 |
| SP50x12 | 12 | 50 | 601 | 3.656 | 6.266 |
| SP100x2 | 2 | 100 | 201 | 4.459 | 5.793 |
| SP100x4 | 4 | 100 | 401 | 4.483 | 6.657 |
| SP100x6 | 6 | 100 | 601 | 4.614 | 7.342 |
| SP100x12 | 12 | 100 | 1201 | 4.801 | 7.751 |

[a] Molecular weight of a star molecule, equals to the sum of that of each arm and one core atom as $M = M_0(fN_{arm} + 1)$, where $M_0 = 1$ is the weight of one monomer.
[b] Radius of gyration $R_g$ of each arm in the bulk (unit: $\sigma$).
[c] Radius of gyration $R_g$ of a whole star molecule in the bulk (unit: $\sigma$).

Cylindrical nanopores with length $L = 60\,\sigma$ and different radii, $R = 4\,\sigma, 7\,\sigma$, and $10\,\sigma$, were modeled by one layer of wall atoms respectively. The distance between two neighbor wall beads was $1.08\,\sigma$, and the interaction parameters of LJ potential between all wall beads were set to $\varepsilon_{LJ} = 1.0\,k_B T$ and $\sigma = 0.8$. Wall atoms vibrate around their lattice position under a moderate harmonic potential $\frac{1}{2}Kr^2$, where $K$ is a spring constant that we set to $300\,k_B T/\sigma^2$, and $r$ is the displacement of the atom from its current position to its lattice position. These choices are sufficient to prevent polymer chains from penetrating the wall.[22] The degree of confinement of both the arms and the entire star molecules in the different capillaries are presented in Table S I.

A $40\,\sigma \times 40\,\sigma$ square reservoir of star polymer melts with periodic boundaries perpendicular to the tube axis reside under the capillary. The interaction parameter between polymer chains and wall atoms was set to $\varepsilon_{LJ} = 1.6\,k_B T$ and $\sigma = 1.0$. Initially, the reservoir was sealed to prevent polymers from moving into the capillary. Following equilibration of the polymer melt, the beads in the central region connecting the capillary to the reservoir are removed, and the imbibition process begins. The height of the polymer during imbibition is determined by fitting the density profile to the following function:

$$\rho(h) = \frac{\rho_p + \rho_v}{2} - \frac{\rho_p + \rho_v}{2}\tanh\left[\frac{2(h - h_0)}{l}\right] \quad (4)$$

where $\rho_p$ is the density of polymer melt in the capillary, $\rho_v \approx 0$ is the vapor density in the capillary due to the low polymer vapor pressure, $h_0$ represents the height of interface between polymer melt and its vapor, and $l$ stands for the thickness of the interface. The relationship between the height $h$ and time $t$ of the imbibition process was then monitored. The calculation methods for contrasting with theoretical relation $h(t)$ of Lucas-Washburn equation and evaluating the dynamics of arms or of the whole star are as follows. More calculation details have been given elsewhere.[21]

**Surface tension.** We employed the Irving-Kirkwood's expression[37] to calculate the surface tension of the polymer melt. All data were derived from a flat gas-liquid interface by integrating the asymmetric part of the pressure tensor.

$$\gamma = \int_{-\lambda}^{\lambda} [P_N(z) - P_T(z)]\, dz \tag{5}$$

Here, $2\lambda$ is the thickness of whole gas-liquid interface determined by the variation of $P_N(z) - P_T(z)$, $P_N(z)$ is the normal component and $P_T(z)$ is the tangential component to the interface of the pressure tensor. This integral should be zero both in the gas phase and in the bulk.

**Viscosity.** The viscosity of the polymer melt was calculated by creating a Poiseuille flow.[38] It was obtained by setting up two plates being parallel to the $xy$-plate. Subsequently, the polymer melt was filled between the plates. The density of the filled melt was evaluated from the density profile of the reservoir before imbibition. Then, a body force along $x$-axis was applied on every bead. Following the steady state, we fit

$$v_x(z) = \frac{\rho g_x}{2\eta}(C + Dz - z^2) \tag{6}$$

to the velocity profile along $z$-axis, where $v_x$ is the velocity being parallel to the force, $\rho$ is the density, $g_x$ is the body force acting on the polymer, $\eta$ is the viscosity of the polymer melt, $C$ is a constant (that equals to zero when there is no-slip boundary condition), and $D$ is the distance between two plates. Different values of $g_x$ in the range from $0.2\ k_B T/\sigma$ to $0.6\ k_B T/\sigma$ were employed to validate the result.

**Contact angle.** The contact angle of the star polymer melt on the capillary wall was calculated by placing the same polymer droplet on a plate that has the same lattice structure and interaction parameters with the capillary. After the wetting process reaches equilibrium, we fitted the shape of the droplet and a series of contact angles were obtained.

**Entanglements.** The average number of entanglement points per arm, $Z$, in the system of $N_{arm} = 100$ was calculated by the primitive path analysis algorithm proposed by Everaers et al.[39,40] To begin with, the end atom of each arm and the core atom are fixed in space. Next, the intrachain excluded volume interactions are canceled, while retaining the interchain excluded-volume interactions. Finally, the system was slowly brought to $T = 0$ to minimize its energy. Topology files used in the calculation are saved during capillary filling and the snapshots after operation are shown in Fig. S3. The entanglement length, $N_e$, is then calculated by employing

$$N_e = (N-1)\left(\frac{<L_{pp}^2>}{<R_{pp}^2>} - 1\right)^{-1} \tag{7}$$

where $L_{pp}$ is the contour length, and $R_{pp}$ is the mean-square end-to-end distance. The number of entanglement points $Z$ is then calculated as $Z = N/N_e$.

**Radius of gyration.** The average radius of gyration $R_g$ of an arm or a star molecule were calculated by

$$R_g^2 = \frac{1}{n}\sum_{i=1}^{n}[(x_i - x_{cm})^2 + (y_i - y_{cm})^2 + (z_i - z_{cm})^2] \tag{8}$$

where $n$ is the number of segments of an arm or a star molecule, the subscript $i$ stands for the $i$th bead and $cm$ stands for the center of mass of chains. Components of $R_g$ are also collected by separating the tensor of $R_g$ during imbibition,

$$R_{gz}^2 = \frac{1}{n}\sum_{i=1}^{n}(z_i - z_{cm})^2, R_{gr}^2 = \frac{1}{n}\sum_{i=1}^{n}\frac{1}{2}[(x_i - x_{cm})^2 + (y_i - y_{cm})^2] \tag{9}$$

where $R_{gz}$ is the component of $R_g$ along the direction of capillary filling and $R_{gr}$ is that in the radial direction.

**Configurations.** Polymer chains adsorbed to one interface have three kinds of configurations: loop, train, and tail. We defined the region within $1\,\sigma$ of the capillary wall as the adsorption layer. Chains with no segments in this region are viewed as free chains. The density of each state in the capillary can be calculated by $n_{conf}M_0/V_{capillary}$, where $n_{conf}$ is the sum of the number of segments in one certain configuration, and $V_{capillary}$ is the volume of the capillary. The data were collected when the wetting height was twice that of the $R_g$ of arms, as $h > 10\,\sigma$.

**Self-correlation function of the core-to-end vector.** After full imbibition, the self-correlation function of the core-to-end vector of the arms was calculated by

$$C_n(t) = \frac{1}{n}\sum_{j=1}^{n} \frac{\langle \boldsymbol{r}_j(t) \cdot \boldsymbol{r}_j(0) \rangle}{\langle |\boldsymbol{r}_j(0)|^2 \rangle} \quad (10)$$

where $n$ is the total number of arms in the capillary, and the vector $\boldsymbol{r}_j$ denotes the difference between the position of the end atom and core atom of $j$th arm.

**Mean square displacement of core segments.** The mean square displacement of core segments under confinement was calculated by

$$MSD_z(t) = \frac{1}{n}\sum_{i=1}^{n}(z_{it} - z_{i0})^2$$

$$MSD_r(t) = \frac{1}{n}\sum_{i=1}^{n}\frac{1}{2}[(x_{it} - x_{i0})^2 + (y_{it} - y_{i0})^2] \quad (11)$$

where $MSD_z$ is the displacement along the axis of capillary, $MSD_r$ is the displacement in the radial direction, $n$ is the total number of core atoms in the capillary, the subscript $i0$ stands for the position of $i$th core atom at initial time, and $it$ stands for the position of $i$th core atom at time $t$.

All simulations were performed using LAMMPS[41] and carried out under a constant temperature maintained by DPD thermostat[42] with friction parameter $\xi = 0.5\sqrt{mk_BT/\sigma^2}$ and interaction cutoff $r_{cut} = 2.5\,\sigma$. The friction parameter $\xi$ determines the strength of the dissipative force and reflects the efficiency of keeping the temperature constant following thermal disturbance. We used the velocity-Verlet algorithm[43] with a time step $\Delta t = 0.005\ \tau_{LJ}$ for integration, where $\tau_{LJ} = \sqrt{m\sigma^2/\varepsilon_{LJ}}$ is the standard LJ-time.

**III. RESULTS AND DISCUSSION**

**A. Deviated imbibition dynamics from prediction of Lucas-Washburn equation**

To calculate the theoretical value of LWE prefactor, the data of surface tension $\gamma$, viscosity $\eta$, and contact angle $\theta$ for the star polymer melts were collected and plotted in Fig. 1. For linear chains, the variation of suface tension with molecular weight follows the relation $\gamma = \gamma_\infty - kM^{-1}$, where $\gamma_\infty$ is the surface tension at infinite molecular weight and $k$ is a constant determined by the nature of polymer.[44-46] Figure 1 (b) shows that the surface tension of star polymers exhibits a linear dependence on $M^{-1}$ with functionality-dependent slopes, while the slope $k$ increases with functionality. This further justifies the notion that despite the very different chain topology, the major factor governing the surface tension is the presence of chain ends.[33] The variation of surface tension with $N_{arm}$ is shown in Fig. 1 (a), where the solid lines represent the fitting results using $\gamma = \gamma_\infty - KN_{arm}^{-1}$. The good fit also reveals that there should be a relation $KN_{arm}^{-1} = kM^{-1}$. Since $M \approx fN_{arm}$, we can get $K = k/f$.

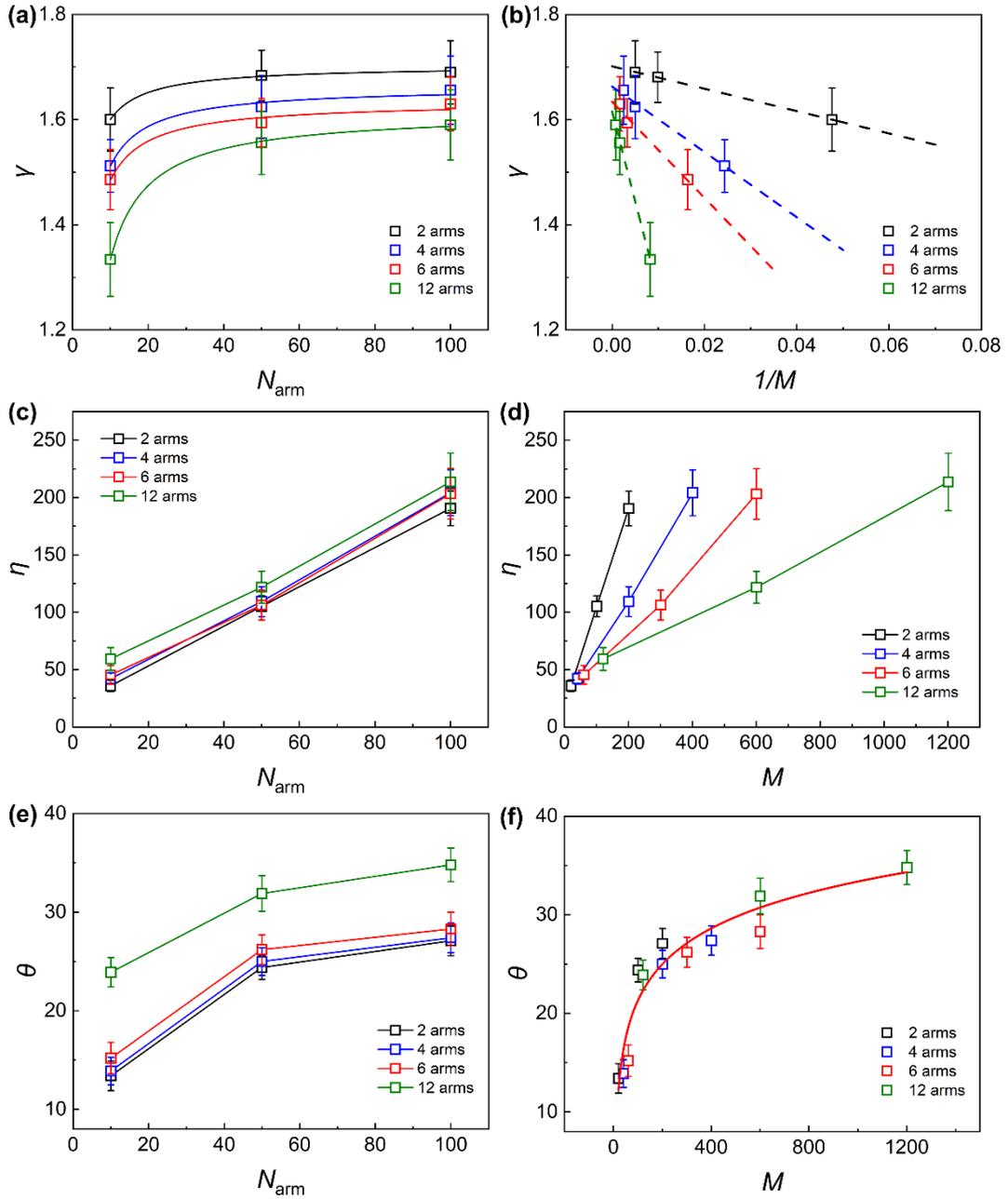

**FIG. 1.** Basic properties of different star polymer melts ($f = 2, 4, 6,$ and $12$) required by LWE. (a) Surface tension $\gamma$ as a function of the arm length, $N_{arm}$, and (b) of the molecular weight $M$. (c) Viscosity $\eta$ in the bulk as a function of the length of arm $N_{arm}$ and (d) of the molecular weight $M$. (e) Contact angle $\theta$ on a plate that has the same characteristics with the capillary, as a function of the length of arm $N_{arm}$ and (f) of the molecular weight $M$.

Figure 1 (c) shows the increasing intrinsic viscosity $\eta$ of star polymers with $N_{arm}$, or with $f$ when $N_{arm}$ is kept constant. Figure 1 (d) show that $\eta$ decreases with $f$ when $M$ is kept constant, which means $N_{arm}$ is reduced. Both results confirm that the change in arm length determines the viscosity of star polymers.[34] Due to the computational limitation, the longest arm in our simulation is of a moderate level of entanglements[21], the scaling relation follows $\eta \propto N_{arm}$ and $\eta \propto M$ (instead of exponential).

Figure 1 (e) illustrates that when the solid-liquid interaction keeps constant, the contact angle $\theta$ of star polymer melts increases with $N_{arm}$ when $f$ is kept constant. $\theta$ also increases with $f$ for any given $N_{arm}$, which might be attributed to an increasing stiffness of chains.[27,47] Figure 1 (f) shows that $\theta$ increases with $M$ and the variation seems functionality-independent since the fitting results can overlap roughly within the margin of error. This further suggests that when discussing $\theta$ of star polymers, the parameters $f$ and $N_{arm}$ should be combined into $M$.

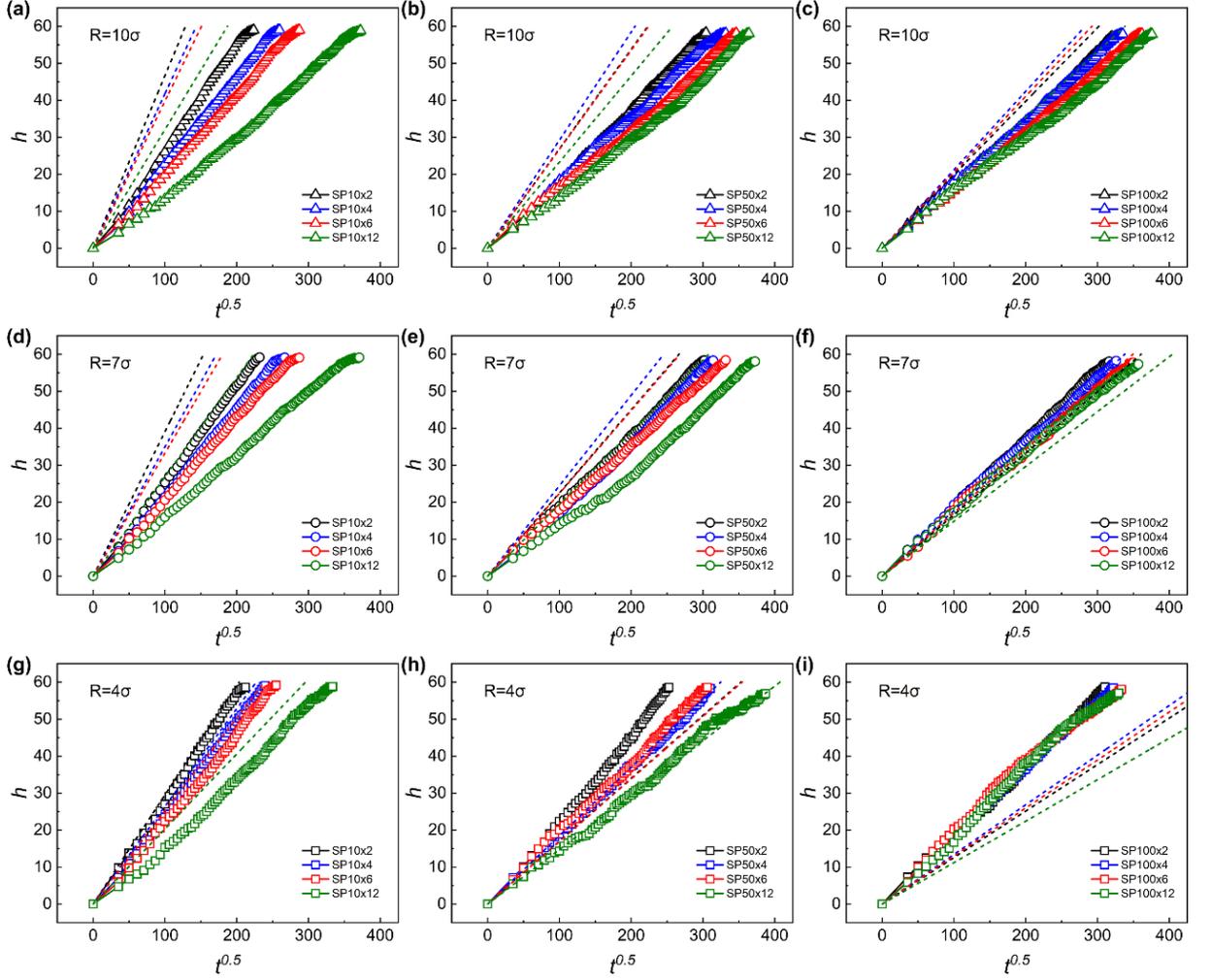

**FIG. 2.** Comparison of actual capillary filling with the theoretical LWE predictions for star polymer melts in nanopores with different radii, (a) ~ (c) $R = 10\,\sigma$, (d) ~ (f) $R = 7\,\sigma$, and (g) ~ (i) $R = 4\,\sigma$. The dashed lines represent the theoretical prediction of LWE, and the solid lines stand for the real imbibition processes as obtained in the simulation. Star polymer melts of different functionalities, $f = 2, 4, 6$ and $12$, are represented by black lines, blue lines, red lines and green lines, respectively.

The theoretically predicted relations of $h(t)$ in different systems were calculated according to the data in Fig. 1 and shown as the dashed lines in Fig. 2. The actual data points for $h(t)$ according to Eq. (4) are also plotted and further fitted by solid lines. The results show that both the actual and theoretical wetting speed decrease with functionality $f$ when $N_{arm}$ is fixed. This can be attributed to the increase of intrinsic viscosity $\eta$ with $f$. More significantly, with the star architecture ($f > 2$) a non-monotonic deviation of the imbibition dynamics from the theoretical LWE is observed as with the linear chains ($f = 2$).[21] When the length of arm is short (e.g., $N_{arm} = 10$), the imbibition process is slower than theoretically predicted. When the arm length is longer (e.g. $N_{arm} = 100$), the dynamics reverses and the filling process is faster than theoretically predicted. In previous work, a theoretical model has been established to explain the breakdown of LWE for linear polymer systems.[18] A dead zone of adsorbed polymers (i.e. reduced effective pore radius) combined with decreased free energy of the polymer under confinement slows down the imbibition process. For long-chain systems, reptation-like plug flow leads to enhanced microscopic flow of entangled chains and faster imbibition.[20,21] For star polymers, because the arms can be viewed as linear chains whose one end is spatially restricted at the star center, the mechanism behind the reversal of imbibition dynamics bears similarities with that of linear polymers (i.e., the "dead zone" effect, the reptation effect, etc.). The functionality-dependent spatial restriction effect can modulate some dynamic features, such as adsorption and entanglements. This thereby influences those acceleration or deceleration effects of filling, which we will discuss in the next sections.

Considering the critical influence of dynamic contact angle $\theta_d$ on simple liquid[48,49], we also calculated the $\theta_d$ as a function of wetting height $h$, as shown in Fig. S1. The results indicate that at the beginning of imbibition ($h < 10$), $\theta_d$ decreases rapidly to a constant value slightly higher than the static contact angle $\theta$. The difference between stable $\theta_d$ and $\theta$ increases as the arm length $N_{arm}$ of system decreases, but does not exceed 20° at most. We further verify the effect of $\theta_d$ by comparing the actual imbibition dynamics with the prediction of the LW equation modified by $\theta_d$, as shown in Fig. S2. The results show that the actual imbibition still deviates significantly from the theoretical prediction, which illustrates the use of $\theta_d$ for modifying LW equation has very limited effect for polymer system.

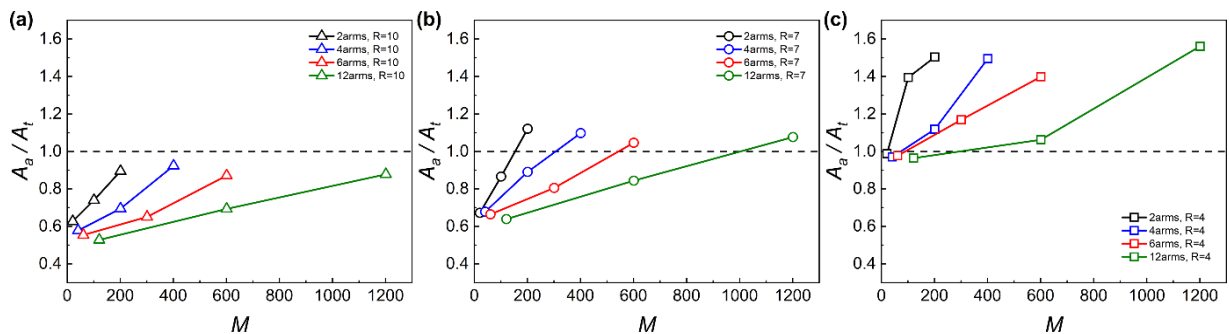

**FIG. 3.** Variation of the ratio of actual prefactor $A_a$ and theoretical prefactor $A_t$ with the molecular weight $M$ of star polymers of different functionality ($f = 2, 4, 6$ and $12$), in capillaries with different radii (a) $R = 4\,\sigma$, (b) $R = 7\,\sigma$, and (c) $R = 10\,\sigma$.

To show the trend clearer, we plot the ratios of actual prefactors $A_a$ and theoretical prefactors $A_t$ at different molecular weights $M$ of the star polymer in Fig, 3. Interestingly, when $M$ is kept constant, reducing the functionality $f$ helps trigger the reversal. This finding suggests that it is highly beneficial to design star polymer nano-fillers with fewer arms when they need to fill pores in a short amount of time. For example, this reduced-arms strategy may help grouting coating of star polymers to penetrate deeper into thin holes that need to be repaired, bypassing the complete curing of the coating. Additionally, Figure 3 shows that the reversal of imbibition dynamics is strongly depending on the degree of confinement (being faster for highly confined polymers), similar to linear polymers. This further verifies the same physics behind the reversal in imbibition kinetics despite the different chain topology.

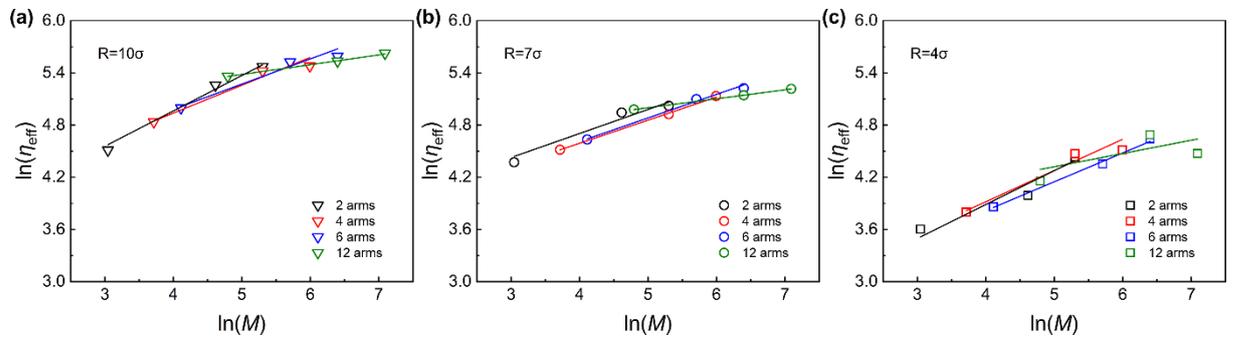

**FIG. 4.** Variation of the effective viscosity $\eta_{eff}$ with the molecular weight $M$ of star polymers of different functionality ($f = 2, 4, 6$ and $12$), in capillaries with different radii (a) $R = 4\,\sigma$, (b) $R = 7\,\sigma$, and (c) $R = 10\,\sigma$.

In order to show the acceleration and deceleration effect of imbibition kinetics in a comprehensive way, we calculated the effective viscosity $\eta_{eff}$ of star polymer melts from the LWE as $\eta_{eff} = \gamma R\cos\theta/2A_a^2$, as shown in Fig. 4. The relation between $\eta_{eff}$ and $M$ follows the form $\eta_{eff} \propto M^\alpha$, where $\alpha$ has shown to depend on the degree of entanglement and confinement in linear polymers.[21] By taking the logarithm and further applying a linear fit, we obtained the value of $\alpha$, which is reflected as the slope of fitting line in Fig. 4. The result illustrates that the polymer architecture is also an important factor influencing the value of $\alpha$. For symmetric star polymers, $\alpha$ was found to decrease with functionality. Furthermore, the effect of the star topology is more pronounced than the change in degree of confinement in our model, as shown in Fig. 4. Thus, we take the average $\alpha$ in capillaries of three radii, where $\alpha = 0.39, 0.33, 0.28$ and $0.12$, for $f = 2, 4, 6$ and $12$, respectively. It shows a good linear relation between $\alpha$ and $f$. In addition, from Fig. 4, the lower $\alpha$ in the system of higher functionality $f$ is mainly caused by the higher $\eta_{eff}$ at low $M$ or $N_{arm}$. We deem that it is the stronger adsorption of high functionality star polymers that enhances the effect of the "dead zone" as recently found experimentally.[50]

**B. Chain dynamics during imbibition**

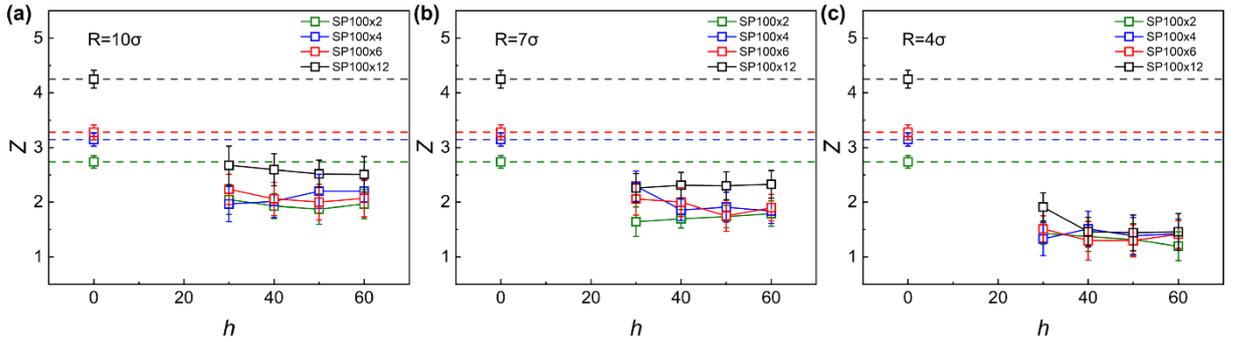

**FIG. 5.** Variation of the average number of entanglement points per arm $Z$ of star polymers of $N_{arm} = 100$ with the wetting height $h$, during imbibition in the cases of (a) $R = 10\ \sigma$, (b) $R = 7\ \sigma$, and (c) $R = 4\ \sigma$.

Understanding the behavior of chains during imbibition into a nanopore not only helps identify the physics behind the breakdown of LWE, but also has practical implications for various processing techniques. According to previous works, disentanglement is a main factor that causes the low effective viscosity during capillary filling and the reversal of the imbibition dynamics.[20,21] Therefore, we first present the average number of entanglement points per arm $Z$ at different wetting heights, $h$, during imbibition for the stars with $N_{arm} = 100$ (obtained from by Eq. (7)), in Fig. 5. In this case entanglements could be formed by two arms from different star polymers, as well as within the same star. The value of $Z$ at $h = 0$ represents the bulk case before the onset of capillary filling. In the bulk, the higher functionality induces a higher degree of entanglement for the arms. This is mainly because, when focusing on the dynamics of arms, each arm can be considered as a linear chain with added spatial constraint at one end. Evidently, topological constraints are intensified by increasing functionality.[51,52] In the process of capillary filling, chains can only move along the "reptation tube" and are oriented due to the confinement and flow. This motional mode obviously results to reduce entanglements, which is more significant under higher degree of confinement, as shown in Fig. 5. Furthermore, the number of remained entanglement points shows a dependence on functionality $f$. The system of higher $f$ is prone to retain more entanglement points. However, this effect gradually weakens with increasing degree of confinement. In the case of $R = 4\,\sigma$, where the degree of confinement on the arms of $N_{arm} = 100$ is larger than 1 (i.e., $R_g^{(arm)} > R$), the number of remained entanglement points of each system is roughly equal.

Of particular importance is understanding the distribution of remained entanglement points within the capillary, as it provides insights into the varying strength of topological constraints at different positions. The primitive path analysis algorithm proposed by Everaers *et al*[39,40] is no doubt a strong tool to obtain the number of entanglement points, but does not suffice at calculating their exact locations. Thus, we can only approximately judge the location of original entanglement points by the topological snapshots, as shown in Fig. S3. The snapshots shows that the remained entanglement points are mainly distributed in the vicinity of the capillary wall and the front of the flow. The former one is caused by extra topological constraints imposed by the adsorbed arms as observed experimentally[50], and the latter one is induced by the radial velocity component at the front of the flow.[22]

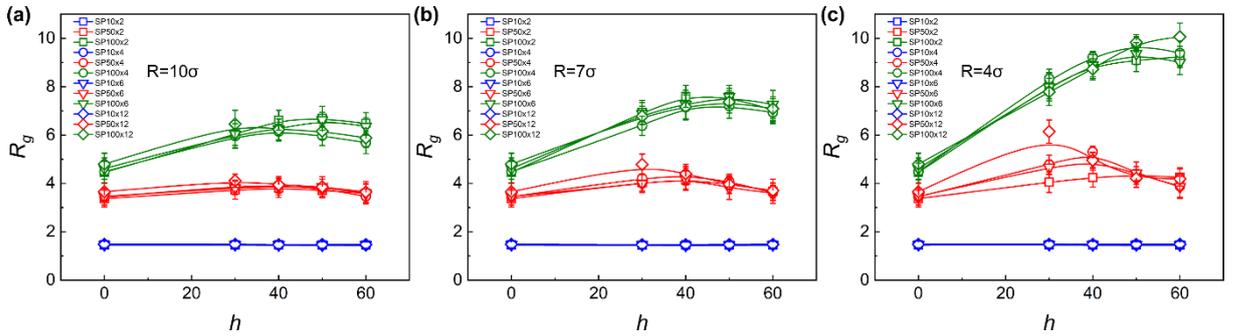

**FIG. 6.** Variation of the average radius of gyration $R_g$ of the arms in the capillary with the wetting height $h$ during the imbibition processes in the cases of (a) $R = 10\ \sigma$, (b) $R = 7\ \sigma$, and (c) $R = 4\ \sigma$.

The orientational stretching due to the special motional mode under confinement is considered as the reason for the disentanglement of chains.[20,21] Therefore, we utilized Eq. (8) to calculate the average radius of gyration $R_g$ of the arms within the nanopores during imbibition, and the result is shown in Fig. 6. The initial value of $R_g$ at $h = 0$ stands for the $R_g$ of arms in the bulk (before the imbibition). In Fig. 6, a clear extension of arms can be observed during capillary filling under confinement for systems with $N_{arm} = 50$, and $100$. For systems with $N_{arm} = 10$, these short arms are difficult to form random coils initially (i.e. short arms are already stretched), so the change in $R_g$ is very slight. As the capillary is close to being completely filled, the effect of flow gradually diminishes, and this extension exhibits a tendency to revert back. Both, the length of the arms and the degree of confinement contribute to an increase in the maximum extension and recovery time. Hence, it can be concluded that flow initiates the extension of chains, while the degree of confinement determines the extent of extension, which subsequently influences the degree of disentanglement. However, it is difficult to identify a clear functionality dependence of the degree of extension for star polymers. The $R_g(h)$ curves of stars bearing the same $N_{arm}$ but different $f$ nearly overlap. It suggests that though the spatial restriction imposed on the end of arms is strengthened as $f$ increases, its impact on the extension of arms during the imbibition process is weak.

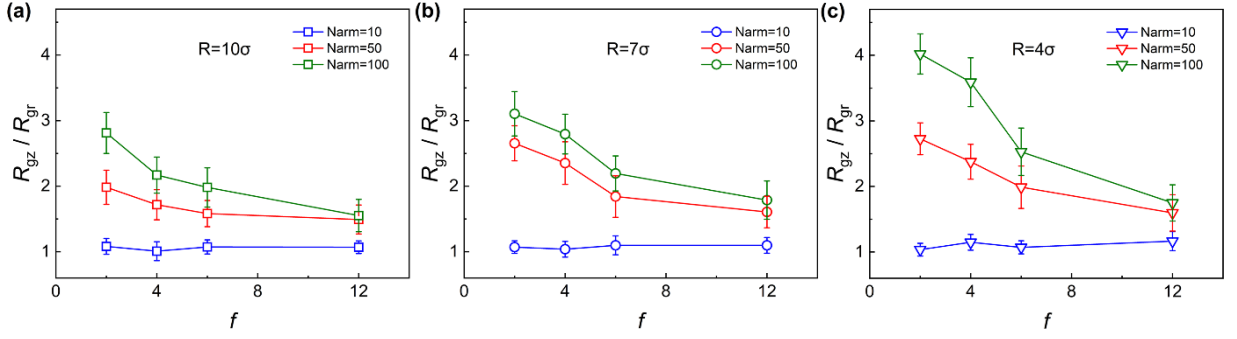

**FIG. 7.** The ratio of the components of $R_g$ of the whole star polymer along the direction of capillary filling $R_{gz}$ and in the radial direction $R_{gr}$ within capillaries with (a) $R = 10\,\sigma$, (b) $R = 7\,\sigma$, and (c) $R = 4\,\sigma$. The data are collected when the wetting height is more than twice the $R_g$ of arms in static status. The curves reflect the relation between the change in shape of star molecules and their functionality $f$, during nanopore filling.

Besides the average radius of gyration $R_g$ of the arms, we also calculated $R_g$ of the whole stars within the nanopores. In order to demonstrate the change in shape of the stars during imbibition, we separated the components of $R_g$ along the direction of capillary filling ($R_{gz}$) and in the radial direction ($R_{gr}$), by Eq. (9). The ratio of $R_{gz}$ and $R_{gr}$ describes the shape of the star polymers. Generally, the value of $R_{gz}/R_{gr}$ of coil-like chains in the static bulk is around 1, especially for the star polymers. Figure 7 reveals that the shape of star molecules undergoes a transformation from spherical to ellipsoidal during capillary filling, as $R_{gz}$ becomes larger than $R_{gr}$. Furthermore, the degree of shape change depends on functionality $f$. Star polymers with larger number of arms are more prone to keep the original spherical shape. This can be attributed to the increased stiffness of star polymers with high functionality $f$.[27] To further prove that, we collected the distribution of both the core atoms of star molecules and end atoms of the arms during imbibition, as shown in Fig. S4 and Fig. S5. Figure S4 shows that the core atoms in the systems of $f \leq 4$ can still be adsorbed on the capillary wall, as they are distributed over the entire range from $r = 0$ to $r = R$. For the stars with $f = 6$ and $f = 12$, the core atoms can only be found outside a certain distance from the wall, as from $r = 0$ to $r < R$. In addition, this distance increases with functionality $f$. This observation suggests the presence of a rigid region near the core segments due to the star architecture, and this region expands with higher functionality $f$. Once $f$ surpasses a certain threshold, the core segments are no longer able to be adsorbed on the capillary wall. Consequently, the configurational entropy contributed by the core atoms adsorbed on the wall decreases. Figure S5 shows that the end atoms of all systems have a uniform distribution over the entire range from $r = 0$ to $r = R$. This indicates that the adsorption of end atoms is less influenced by chain topology.

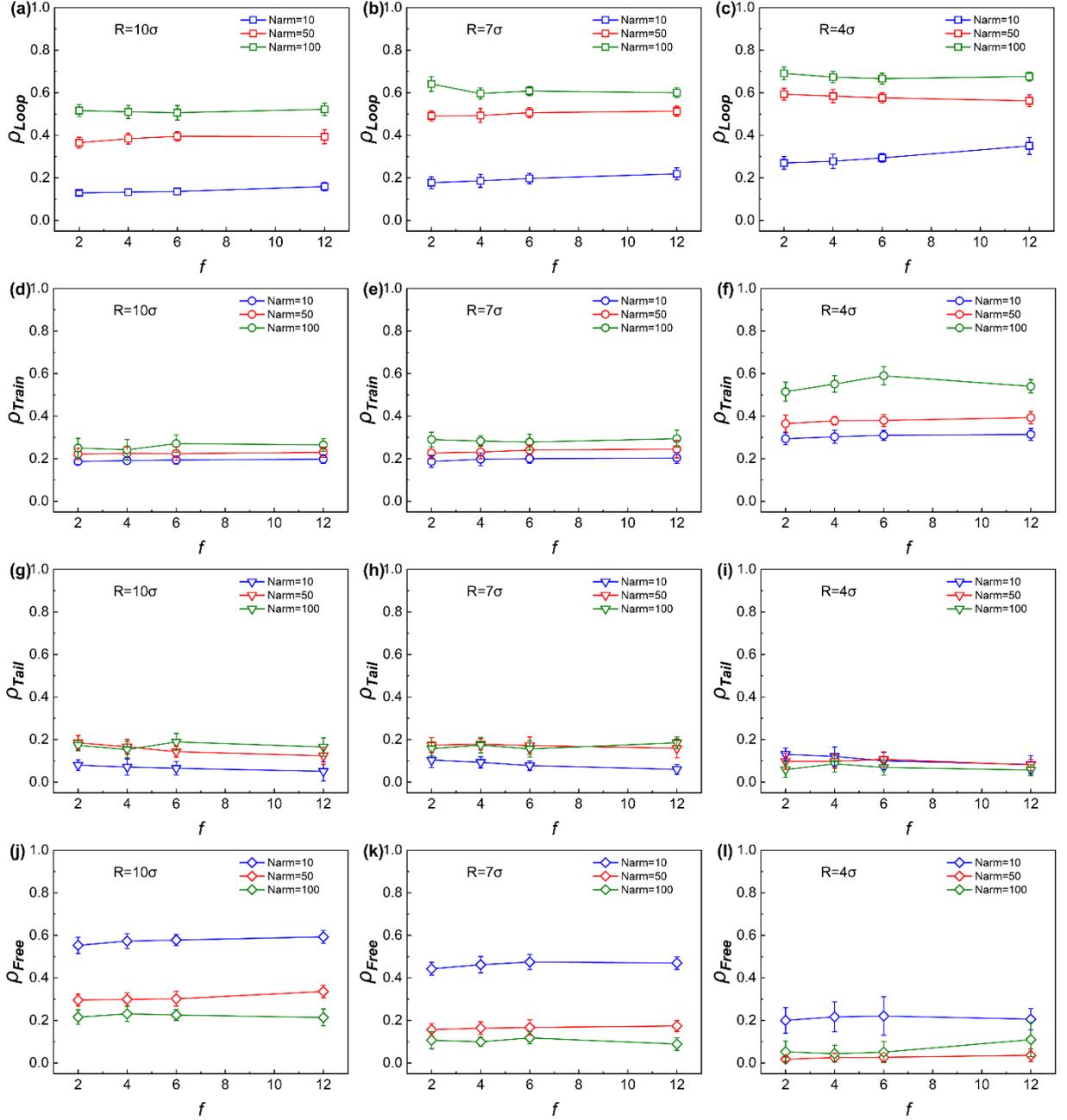

**FIG. 8.** The final density of adsorbed arms in three types of configurations (a) ~ (c) the density of loop $\rho_{loop}$, (d) ~ (f) the density of train $\rho_{train}$, (g) ~ (i) the density of tail $\rho_{tail}$, and density of free chains $\rho_{free}$, (j) ~ (l) in the capillaries with different radii ($R = 10\,\sigma, R = 7\,\sigma,$ and $R = 4$).

Chains near the capillary wall have to pass through several unfavorable configurations before being completely adsorbed, such as loops and trains.[23,24] This process determines the extremely long adsorption time as experimentally found.[53] In simulation, we calculated the number of monomers in each configurations (loops, trains, tails and free chains) during the whole imbibition process, and then obtained the density of each configuration in the capillary, $\rho_{loop}, \rho_{train}, \rho_{tail}$, and $\rho_{free}$. The development of $\rho_{loop}, \rho_{train}, \rho_{tail}$, and $\rho_{free}$ with the wetting time $t$ is shown in Fig. S6. At the first stage, all the density values increase because more arms are entering the capillary constantly. After a period of wetting, the number of segments in each configuration remained constant. The density of the different configurations then forms a stable ratio. Subsequently, we extracted the final density of all systems and the result is depicted in Fig. 8. For loops and trains, with both ends fixed on the wall, the final density is significantly higher in the systems with longer $N_{arm}$. The increase in the degree of confinement also increases the number of such configurations, since the specific surface area for adsorption increases. In addition, $\rho_{loop}$ and $\rho_{train}$ slightly increases with functionality $f$ when $N_{arm} = 10$, but they no longer depend on $f$ when $N_{arm} = 50$ and $N_{arm} = 100$. It demonstrates that the star topologies with high functionality $f$ makes arms more prone to be adsorbed and form loops or trains. However, this effect is gradually weakened with increasing $N_{arm}$. For tails, the configuration of adsorbed arms is such that only one end is fixed on the wall and another is free. When the capillary radius is comparable to the chain length, it is difficult for arms to form long tails under such extreme confinement. So the final density of tails $\rho_{tail}$ decreases with the degree of confinement and is lower than $\rho_{loop}$ and $\rho_{train}$. For density of free arms, $\rho_{free}$ indirectly reflects the amount of adsorption. Figure 8 illustrates that $\rho_{free}$ decreases both with $N_{arm}$ and the degree of confinement. This demonstrates that systems of higher $N_{arm}$ can induce a thicker adsorption layer in the capillary with the same radius. Additionally, the increasing specific surface area of the wall when the capillary radius decreases also leads to a higher amount of adsorption.

## C. Chain dynamics after full imbibition

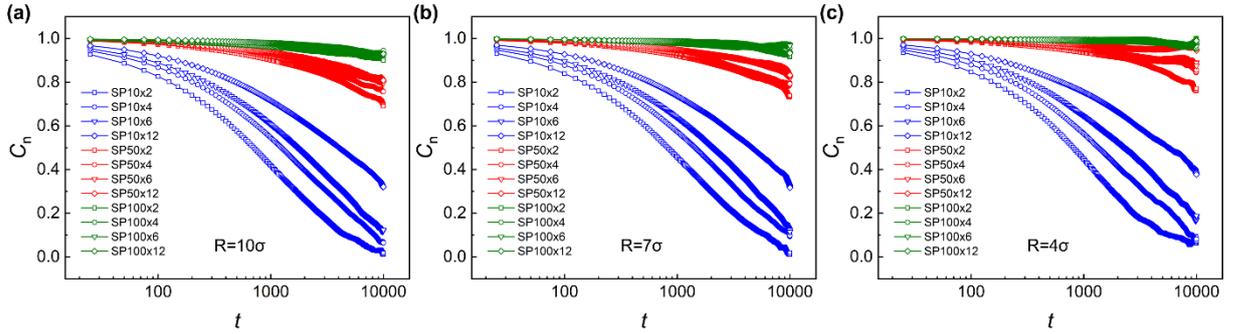

**FIG. 9.** Variation of the self-correlation function of the core-to-end vector of the arms $C_n$ in the capillaries of (a) $R = 10\,\sigma$, (b) $R = 7\,\sigma$, and (c) $R = 4\,\sigma$, during 10000 time units after full imbibition. Lines in different colors represent the systems of $N_{arm} = 10$ (blue), $N_{arm} = 50$ (red), and $N_{arm} = 100$ (green), respectively. Lines with different shapes of points represent the systems of $f = 2$ (square), $f = 4$ (circle), $f = 6$ (triangle), and $f = 12$ (rhombus), respectively.

Following the full imbibition of polymer chains in the capillaries, chains require a period of time to relax towards their equilibrium state under confinement. Recent experiments in type-A polymers confined in cylindrical nanopores have shown very long equilibration times that depend on temperature, pore size, and chain topology.[54] It was shown that star-shaped polymers require longer times to reach equilibrium as compared to linear polymers because of their higher tendency for adsorption.

Therefore, discussing the chain dynamics following imbibition is also relevant in simulations. In order to evaluate the arm relaxation, we calculated the average self-correlation function of the core-to-end vector, as shown in Fig. 9. The data were collected immediately after complete imbibition of the star polymers. Therefore, it well reflects the ability of chains to revert back to the equilibrium state after being disturbed under extreme confinement. The result in Fig. 9 shows that the self-correlation function in the systems of higher $N_{arm}$ decays at a much slower rate, which means long arms require extremely long times to reach equilibrium. This time exceeds the entire wetting time needed to fill the nanopores. Moreover, the self-correlation function also decays more slowly in the system of higher functionality $f$, when $N_{arm}$ of system is kept constant. This is because star polymers rely on the cooperative relaxation of multiple arms to reach an equilibrium state. The higher the functionality $f$ of the star polymers, the longer the process, due to the broadening of the distribution of the relaxation times observed through dielectric spectroscopy.[50] In addition to functionality, the degree of confinement exerts a strong influence on the arm relaxation. To show this, we plot in Fig. S7 the data as a function of capillary radii. The result illustrates that stars with the same $N_{arm}$ and $f$ require a longer relaxation time to reach equilibrium under higher degree of confinement.

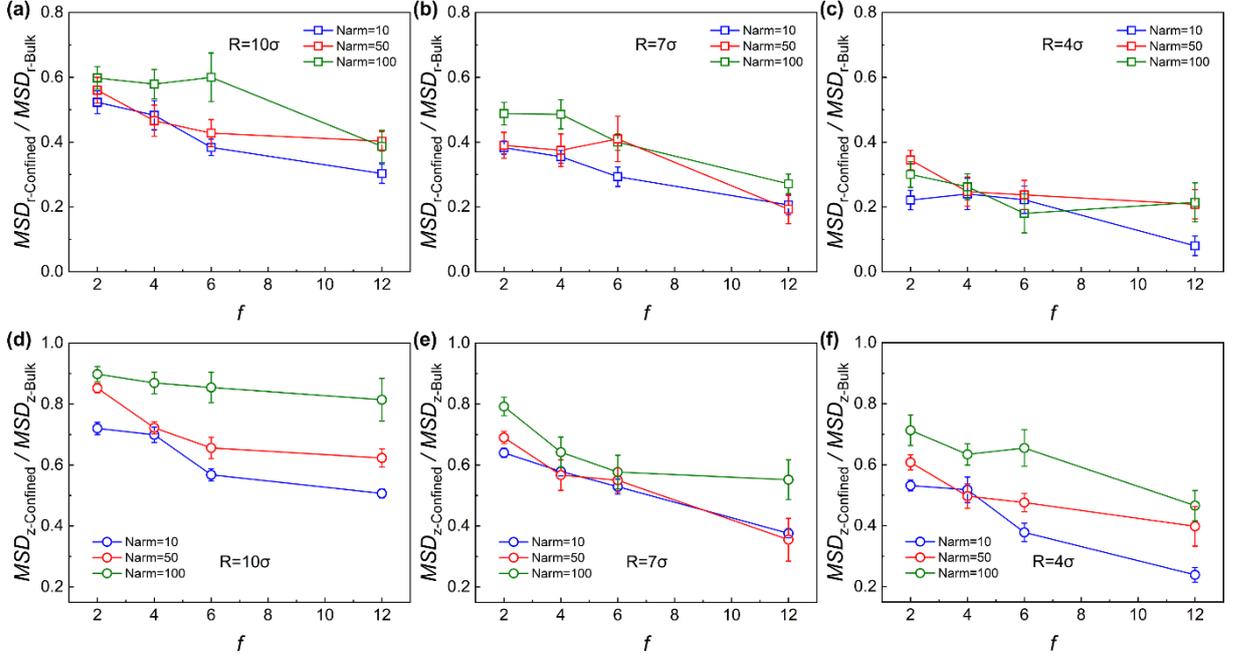

**FIG. 10.** The ratio of the mean square displacement of core segments under confinement and in the bulk, in the cases of (a) ~ (c) in the radial direction, $MSD_{r-Confined}/MSD_{r-Bulk}$ and (d) ~ (f) along the axis of capillary, $MSD_{z-Confined}/MSD_{z-Bulk}$, in the capillaries of $R = 10\,\sigma$, $R = 7\,\sigma$, and $R = 4\,\sigma$, respectively.

As diffusion of molecules is also sensitive to macromolecular topology, we have evaluated the dynamics from the mean-square displacement, $MSD$, of core segments for each star polymer. The evolution of $MSD$ for the core segments in the bulk is depicted in Fig. S8. The result shows that $MSD$ of core segments decreases both with increasing functionality $f$ or $N_{arm}$. This reflects the increasing intermolecular friction of star polymers with increasing $f$ or $N_{arm}$, which further determines the intrinsic viscosity. We also calculated (Fig. S8) the development of *MSD* of core segments in the capillary after arms relax 20000 time steps following the full imbibition. Because of confinement, interfacial friction, and adsorption, both $MSD_{r-Confined}$ and $MSD_{z-Confined}$ decrease below the corresponding bulk value. Interestingly, the degree of reduction is approximately constant, i.e. independent of time. The result is plotted in Fig. 10 as a function of star functionality and $N_{arm}$. It depicts decreasing $MSD_{r-Confined}/MSD_{r-Bulk}$ and $MSD_{z-Confined}/MSD_{z-Bulk}$ for the shorter arms. This reflects an increasing fraction of adsorbed sites with decreasing $N_{arm}$, and the concomitant enhancement of adsorption of core segments. $MSD_{r-Confined}/MSD_{r-Bulk}$ and $MSD_{z-Confined}/MSD_{z-Bulk}$ also decreases with increasing functionality. It further illustrates that star polymers with higher number of arms undergo a stronger effect of adsorption and friction, in agreement with recent experiments.[53] Additionally, the increase in the degree of confinement results to lower $MSD_{r-Confined}/MSD_{r-Bulk}$ and $MSD_{z-Confined}/MSD_{z-Bulk}$, with the effect being more pronounced in the former.

## IV. CONCLUSIONS

In conclusion, the imbibition process of star polymer melts exhibits a reversal in the dynamics of imbibition in nanopores. Systems of shorter arm length $N_{arm}$ penetrate slower than predicted by the LWE, while systems with longer arm length $N_{arm}$ penetrate faster. The physics behind the phenomenon should be same as for linear-chain polymers reflecting the balance between the effects of "dead zone" and disentanglement. A useful strategy to trigger the faster-than-prediction capillary filling of star polymer melts is to decrease the functionality $f$ under a constant molecular weight.

The arms of star polymers with higher $f$ exhibit a higher degree of entanglement in the bulk. Under flow the arms are stretched, causing chains to disentangle. This effect is enhanced with increasing degree of confinement. The remaining entanglements are mainly distributed in the vicinity of the capillary wall and the front of the flow. The star topology induces a stiff region near the core segment which increases with $f$. In addition, systems of higher $f$ experience stronger adsorption, and are more prone to form the configuration of loops and trains. However, this effect is gradually weakened with increasing $N_{arm}$.

After full imbibition, arms of star polymers take a longer time than linear chains to reach the equilibrium state in excellent agreement with recent experiments. Moreover, this time increases with functionality $f$, $N_{arm}$, and degree of confinement. Additionally, the reduced normalized mean square of displacement of core segments following capillary filling, reveals the stronger adsorption and friction in the systems of higher $f$.

**SUPPLEMENTARY MATERIAL**

See the supplementary material for the degree of confinement, dynamic contact angle, snapshots of entanglement, distribution of core atoms and end atoms, configurations in the capillary, self-correlation function of the core-to-end vector of arms, and mean square of displacement of core segments.

ACKNOWLEDGMENTS

This research was supported to Jiajia Zhou by National Key R&D Program of China (2022YFE0103800), the National Natural Science Foundation of China (21774004), the Recruitment Program of Guangdong (2016ZT06C322) and the 111 Project (B18023). Jianwei Zhang acknowledges the funding supported by China Scholarship Council.

AUTHOR DECLARATIONS

**Conflict of Interest**

The authors have no conflicts to disclose.

DATA AVAILABILITY

The data that support the findings of this study are available from the corresponding authors upon reasonable request.

Support Information

# Capillary Filling of Star Polymer Melts in Nanopores

Jianwei Zhang[1], Jinyu Lei[1], Pu Feng[2], George Floudas[3,4,5], Guangzhao Zhang[1,*], and

Jiajia Zhou[6,7,*]

1. Faculty of Materials Science and Engineering, South China University of Technology, Guangzhou 510640, China
2. School of Civil Engineering and Transportation, South China University of Technology, Guangzhou 510640, China
3. Max Planck Institute for Polymer Research, 55128 Mainz, Germany
4. Department of Physics, University of Ioannina, 45110 Ioannina, Greece
5. Institute of Materials Science and Computing, University Research Center of Ioannina (URCI), 45110 Ioannina, Greece
6. South China Advanced Institute for Soft Matter Science and Technology, School of Emergent Soft Matter, South China University of Technology, Guangzhou 510640, China
7. Guangdong Provincial Key Laboratory of Functional and Intelligent Hybrid Materials and Devices, South China University of Technology, Guangzhou 510640, China

**Contents**

1. Degree of confinement
2. Dynamic contact angle
3. Entanglement
4. Distribution of core atoms and end atoms
5. Configurations in the capillary
6. Self-correlation function of the core-to-end vector
7. Mean square displacement of core segments

## 1. Degree of confinement

**TABLE S I.** Degree of Confinement on Arms and Stars in Simulation.

| Star polymers | $R_g^{(arm)}/R_1$ [a] | $R_g^{(arm)}/R_2$ [b] | $R_g^{(arm)}/R_3$ [c] | $R_g^{(star)}/R_1$ | $R_g^{(star)}/R_2$ | $R_g^{(star)}/R_3$ |
|---|---|---|---|---|---|---|
| SP10x2 | 0.15 | 0.21 | 0.36 | 0.22 | 0.31 | 0.55 |
| SP10x4 | 0.15 | 0.21 | 0.36 | 0.26 | 0.37 | 0.64 |
| SP10x6 | 0.15 | 0.21 | 0.37 | 0.28 | 0.39 | 0.69 |
| SP10x12 | 0.15 | 0.21 | 0.37 | 0.31 | 0.44 | 0.77 |
| SP50x2 | 0.34 | 0.48 | 0.84 | 0.45 | 0.65 | 1.13 |
| SP50x4 | 0.34 | 0.49 | 0.86 | 0.54 | 0.77 | 1.34 |
| SP50x6 | 0.35 | 0.50 | 0.87 | 0.55 | 0.79 | 1.38 |
| SP50x12 | 0.37 | 0.52 | 0.91 | 0.63 | 0.90 | 1.57 |
| SP100x2 | 0.45 | 0.64 | 1.11 | 0.58 | 0.83 | 1.45 |
| SP100x4 | 0.45 | 0.64 | 1.12 | 0.67 | 0.95 | 1.66 |
| SP100x6 | 0.46 | 0.66 | 1.15 | 0.73 | 1.05 | 1.84 |
| SP100x12 | 0.48 | 0.69 | 1.20 | 0.78 | 1.11 | 1.94 |

a) $R_1 = 10\ \sigma$. b) $R_2 = 7\ \sigma$. c) $R_3 = 4\ \sigma$.

## 2. Dynamic contact angle

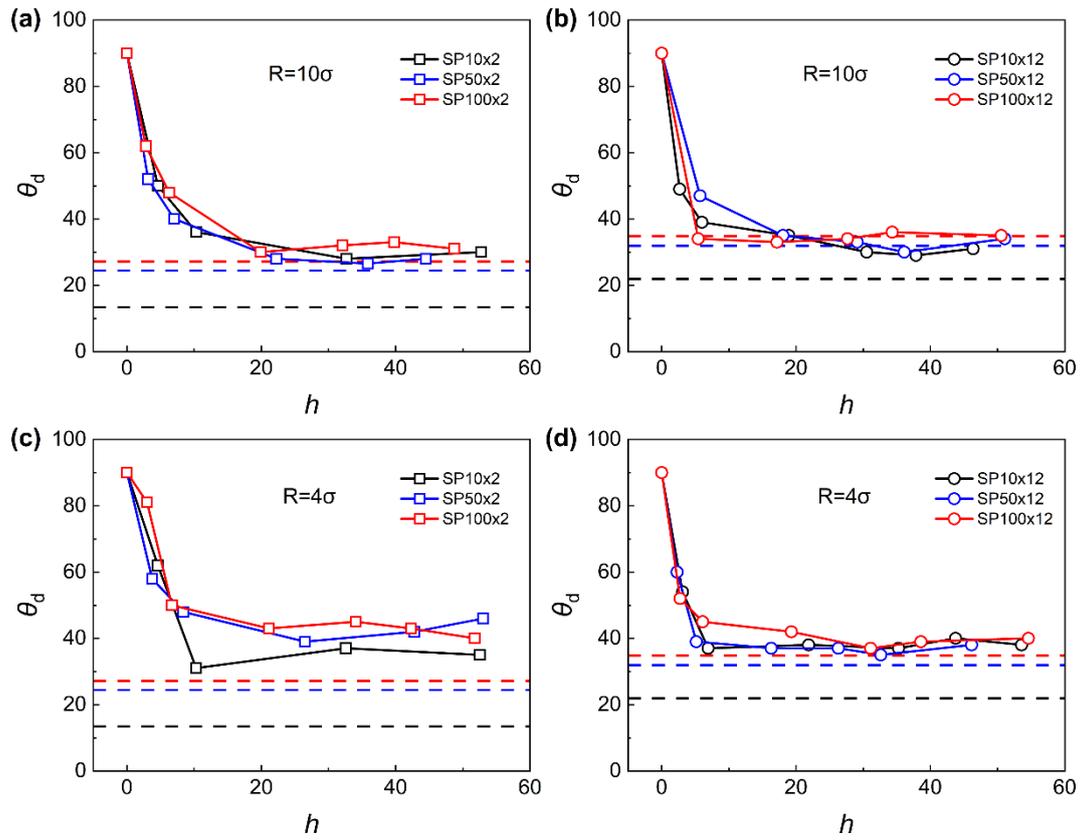

**Fig. S1.** Variation of the dynamic contact angle $\theta_d$ with the wetting height $h$ of both linear polymers and star polymers ($f = 2$ and $12$) with different arm length $N_{arm} = 10$ (black), $N_{arm} = 50$ (blue), and $N_{arm} = 100$ (red), in capillaries with different radii (a)(b) $R = 10\,\sigma$ and (c)(d) $R = 4\,\sigma$.

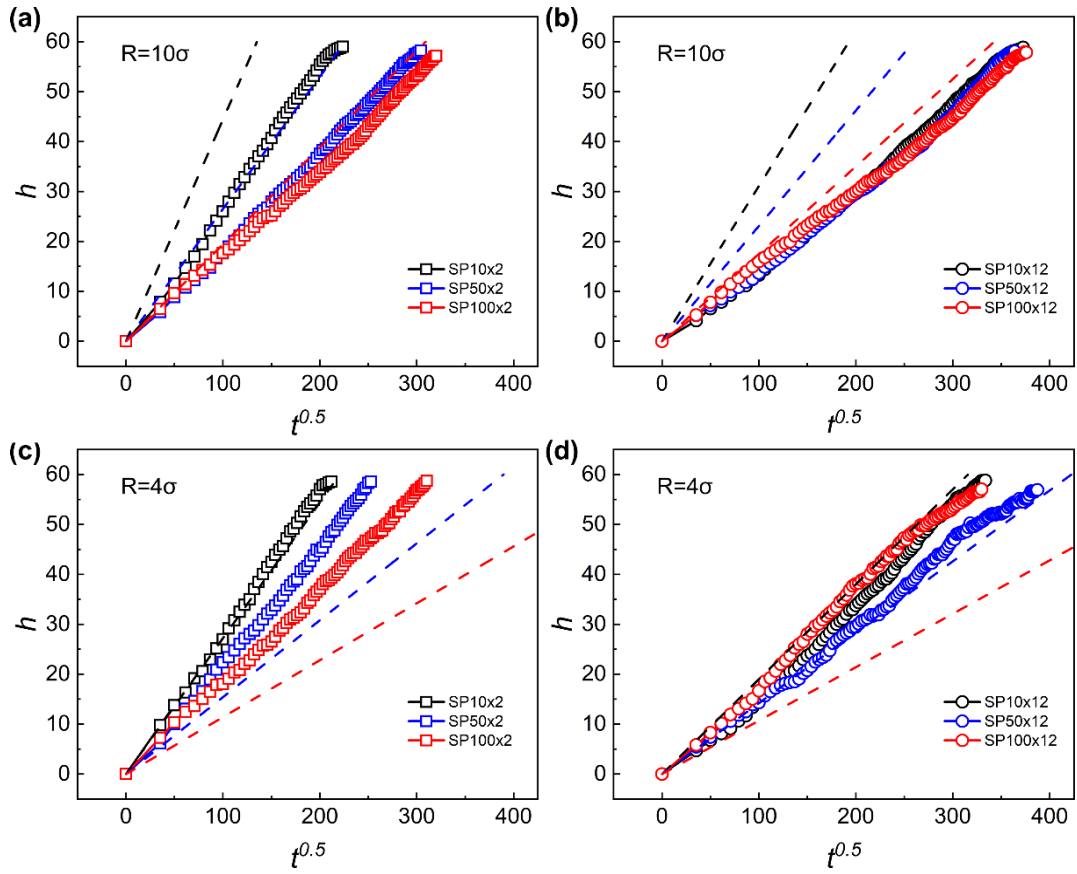

**Fig. S2.** Comparison of actual capillary filling with the theoretical LWE predictions (modified by dynamic contact angle) for both linear polymers and star polymers ($f = 2$ and $12$) with different arm length $N_{arm} = 10$ (black), $N_{arm} = 50$ (blue), and $N_{arm} = 100$ (red), in capillaries with different radii (a)(b) $R = 10\,\sigma$ and (c)(d) $R = 4\,\sigma$.

## 3. Entanglements

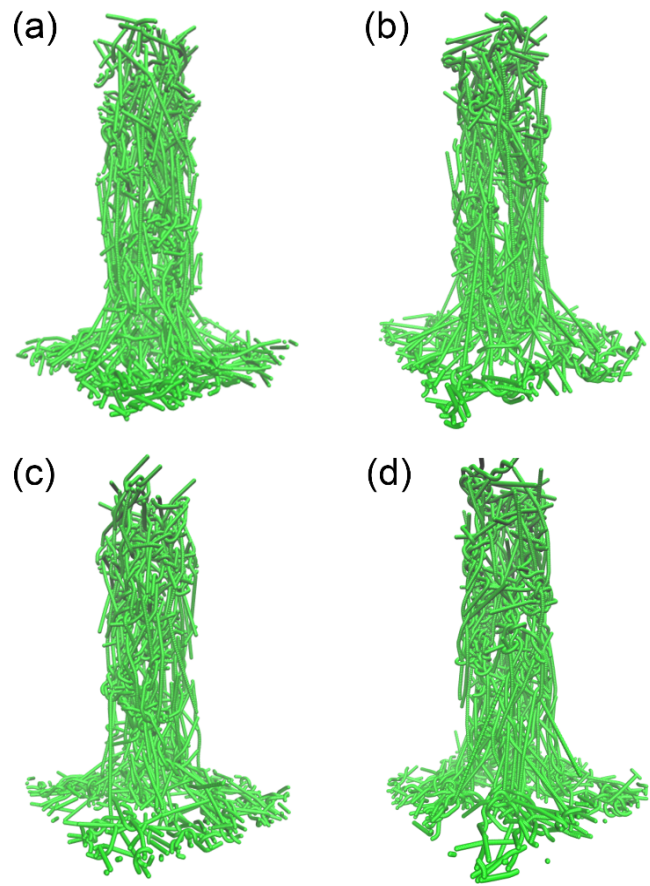

**Fig. S3.** Snapshots of star polymer topologies after the operation of primitive path analysis algorithm, in the cases of (a) SP100x2, (b) SP100x4, (c) SP100x6, and (d) SP100x12 when $h = 60\,\sigma$ during filling in the capillary of $R = 10\,\sigma$.

## 4. Distribution of core atoms and end atoms

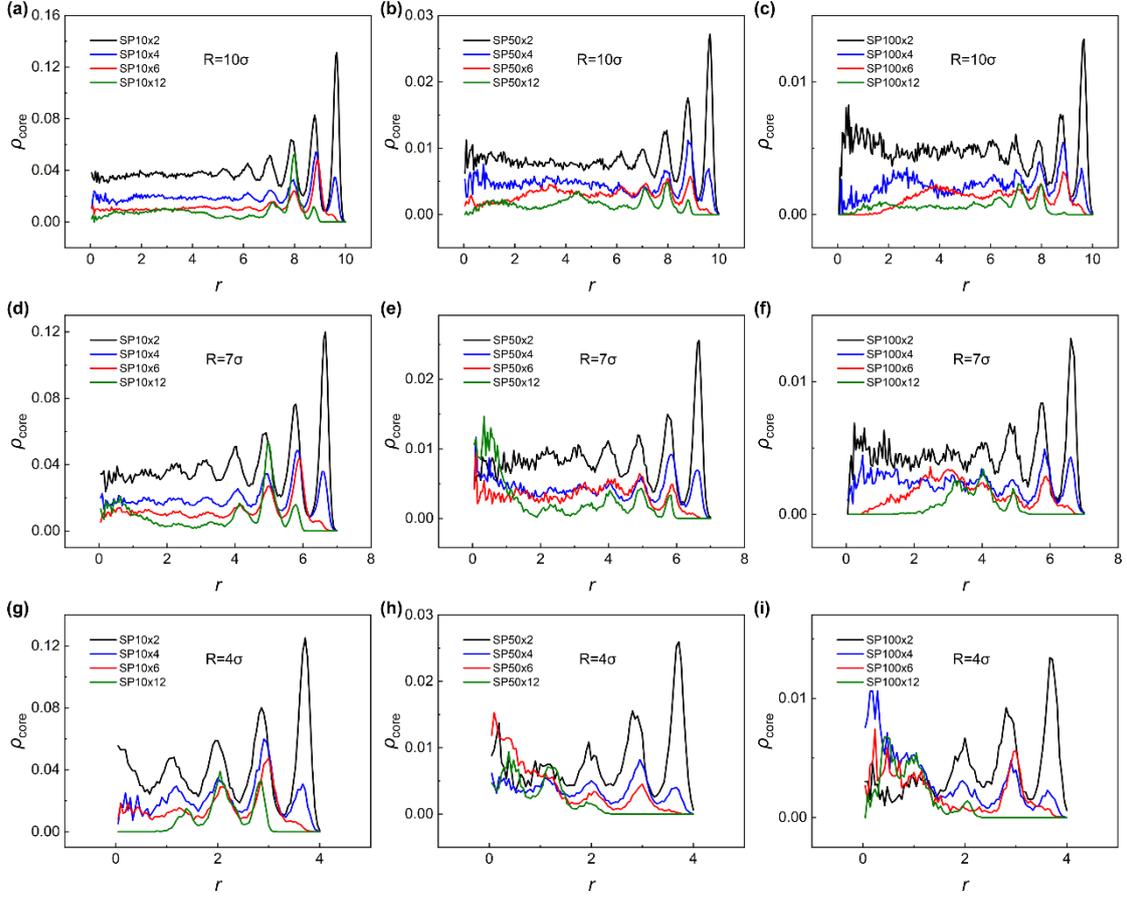

**Fig. S4.** Variation of the density of core atoms $\rho_{core}$ of different star polymer systems within capillary, with the distance to the central axis $r$, in the cases of (a) ~ (c) $R = 10\,\sigma$, (d) ~ (f) $R = 7\,\sigma$, and (g) ~ (i) $R = 4\,\sigma$, during imbibition.

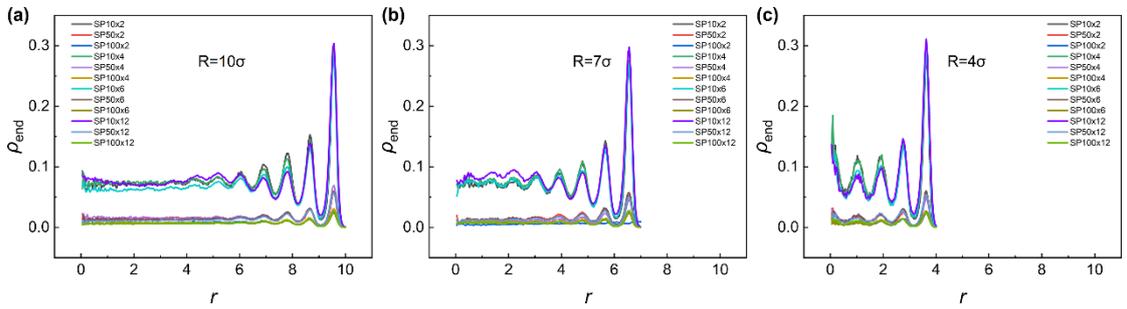

**Fig. S5.** Variation of the density of end atoms $\rho_{end}$ of different star polymer systems within capillary, with the distance to the central axis $r$, in the cases of (a) $R = 10\,\sigma$, (b) $R = 7\,\sigma$, and (c) $R = 4\,\sigma$, during imbibition.

## 5. Chain configurations in the capillary

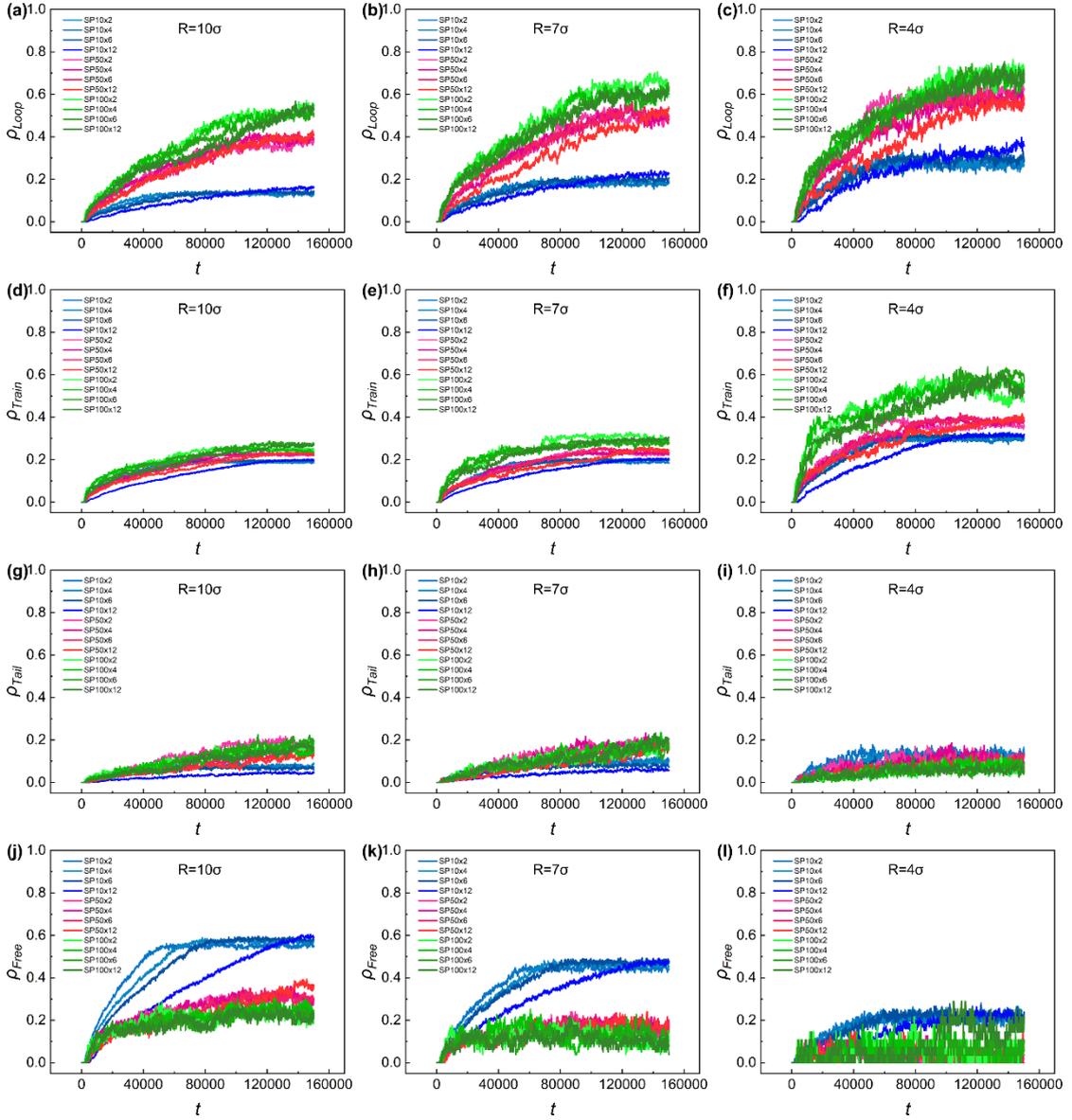

**Fig. S6.** Variation of the density of adsorbed chains in three types of configurations and free chains with the wetting time $t$, in the cases of (a) ~ (c) the density of loop $\rho_{loop}$, (d) ~ (f) the density of train $\rho_{train}$, (g) ~ (i) the density of tail $\rho_{tail}$, and (j) ~ (l) the density of free chains $\rho_{free}$, during imbibition in capillaries of $R = 10\,\sigma$, $R = 7\,\sigma$, and $R = 4\,\sigma$.

## 6. Self-correlation function of the core-to-end vector

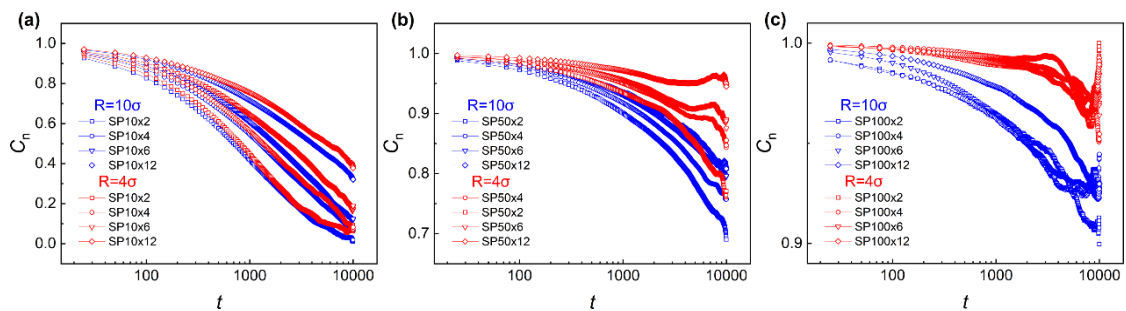

**Fig. S7.** Comparison of the self-correlation function of the core-to-end vector of the arms $C_n$ in the capillaries of $R = 10\,\sigma$ and $R = 4\,\sigma$, in the cases of systems of (a) $N_{arm} = 10$, (b) $N_{arm} = 50$, and (c) $N_{arm} = 100$.

## 7. Mean square displacement of core segments.

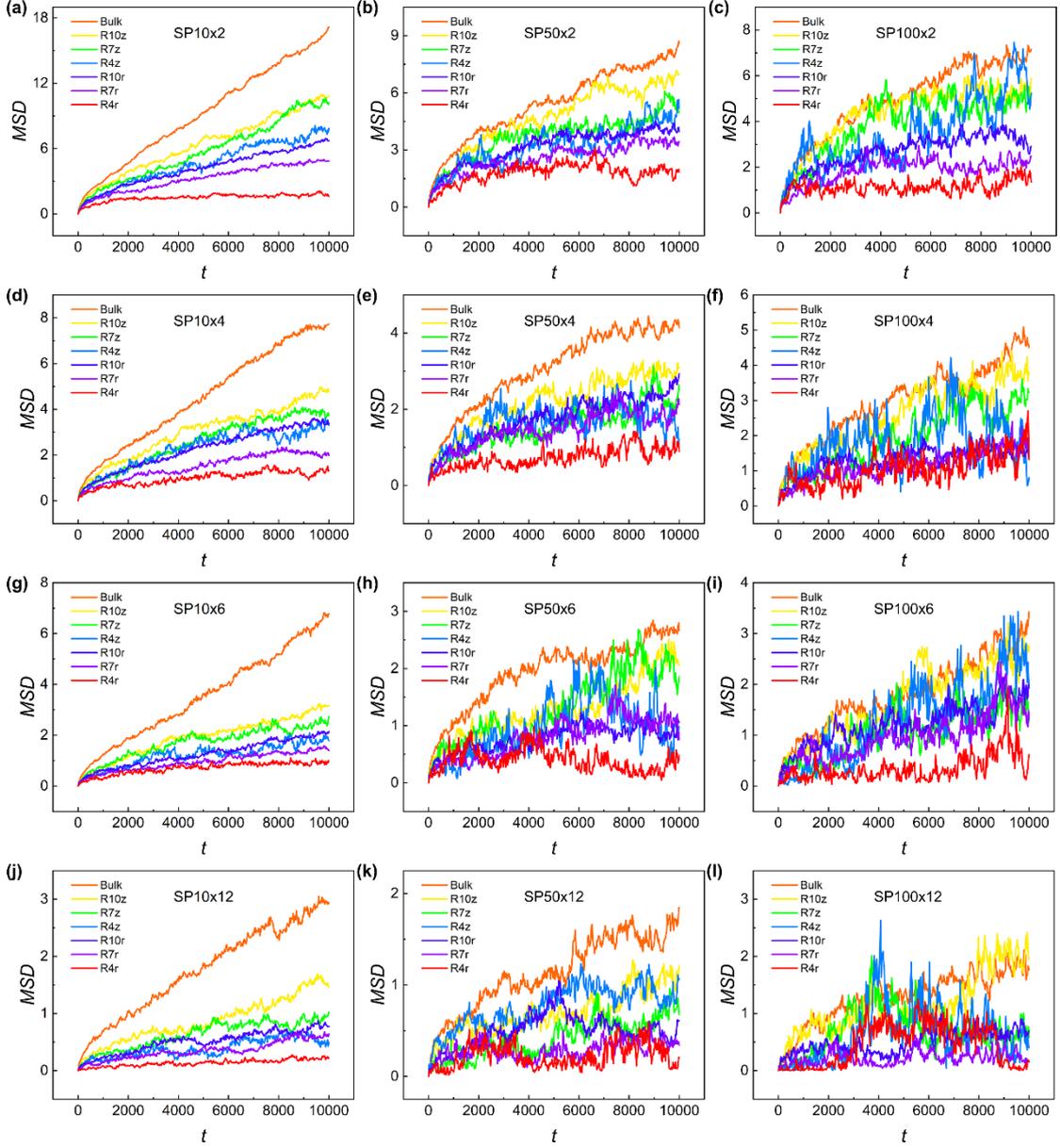

**Fig. S8.** Comparison of the development of the mean square displacement of core atoms in the radial direction and along the axis of capillary, both under confinement and in the static bulk, in the cases of (a) ~ (c) $f = 2$, (d) ~ (f) $f = 4$, (g) ~ (i) $f = 6$, and (j) ~ (l) $f = 12$. Legends represent the radius of capillary and the direction of displacement where the data were collected. For example, "R10z" stands for the case of $MSD$ along the axis of capillary were collected in the nanopore of $R = 10\ \sigma$.

9